\documentclass[aps,pra,twocolumn,showpacs,floatfix,superscriptaddress]{revtex4-1}
\usepackage{color}      
\usepackage{amsmath,amssymb,amsfonts,bbm,graphicx,hyperref,times,subfigure}

\newcommand{\gr}[1]{\boldsymbol{#1}}
\newcommand{\be}{\begin{equation}}
\newcommand{\ee}{\end{equation}}
\newcommand{\bea}{\begin{eqnarray}}
\newcommand{\eea}{\end{eqnarray}}

\newcommand{\sig}{\gr{\sigma}}

\def\Tr{\hbox{Tr}} \def\sigmaCM{\boldsymbol{\sigma}}
\begin{document}
\title{Quantum filtering of a thermal master equation with purified reservoir}
\date{\today}
\author{Marco G. Genoni}
\affiliation{Department of Physics \& Astronomy, University College London, 
Gower Street, London WC1E 6BT, United Kingdom}
\email{marco.genoni@ucl.ac.uk}
\author{Stefano Mancini}
\affiliation{School of Science and Technology, University of Camerino, I-62032 Camerino, Italy \\
and INFN, Sezione di Perugia, I-06123 Perugia, Italy}
\author{Howard M. Wiseman}
\affiliation{Centre for Quantum Computation and Communication Technology (Australian Research Council), Centre for Quantum Dynamics, Griffith University, Brisbane, QLD 4111, Australia}
\author{Alessio Serafini}
\affiliation{Department of Physics \& Astronomy, University College London, 
Gower Street, London WC1E 6BT, United Kingdom}

\begin{abstract}
We consider a system subject to a quantum optical master equation at finite temperature and study a class of conditional dynamics  
obtained by monitoring its totally or partially purified environment.
More specifically, drawing from the notion that the thermal state of the environment may be regarded as the local state of a lossy
and noisy two-mode squeezed state, we consider conditional dynamics (``unravellings'')  
resulting from the homodyne detection of the two modes of such a state. 
Thus, we identify a class of unravellings parametrised by the loss rate suffered by the environmental two-mode state, 
which interpolate between direct detection of the environmental mode alone (occurring for total loss, whereby no correlation between the two 
environmental modes is left) and full access to the purification of the bath (occurring when no loss is acting and the two-mode state of the environment is pure). 
We hence show that, while direct detection of the bath is not able to reach the maximal steady-state squeezing 
allowed by general-dyne unravellings, such optimal values can be obtained when a fully purified bath is accessible. 
More generally we show that, within our framework, any degree of access to the bath purification improves the performance of filtering protocols in terms of achievable squeezing and entanglement.
\end{abstract}
\maketitle
\section{Motivation and summary}

A thorough understanding of the possibilities offered by conditional quantum dynamics, as well as  
a classification of the measurements that enact them, are paramount to the 
design of protocols for the coherent control of open quantum systems.
Over the last thirty years, much progress has been made towards a comprehensive description of the quantum filtering of
systems described by continuous, canonical degrees of freedom \cite{WisemanMilburn,haus86,shapiro87,Belavkin,wiseman93,wiseman94a,wiseman94b,doherty99,doherty00,steck04,ahn02,serafozzireview}. 
In particular, a very general theoretical framework 
has been identified encompassing all diffusive conditional dynamics on such systems, 
that is all conditional dynamics that can be described by multivariate 
quantum Wiener processes \cite{wiseman01,wiseman05,ChiaWise}.
Such dynamics are conditioned by continuous general-dyne detections of the environment. 
General-dyne detection schemes form a class that can be implemented by adding ancillary modes 
in a Gaussian state, applying a Gaussian unitary transformation on the dilated system,
and then performing any possible homodyne detection on it
\cite{ChiaWise,GeneralDino}. They are hence experimentally viable and include the well known homodyne and heterodyne detection schemes. Let us recall the reader that, in the quantum optical literature, the term ``homodyne'' detection refers to the projective measurement on the eigenbasis of a canonical degree of freedom, such as the spectral measurement associated to the position or momentum operators $\hat{x}$ or $\hat{p}$. 

For systems with quadratic (in $\hat{x}$ and $\hat{p}$) Hamiltonians and linear couplings to the environment, 
filtering through general-dyne detection preserves the Gaussian character of the system's state, and allows for the analytical treatment and optimisation of several figures of merit, including squeezing and quantum correlations (if quantified by a measure that is computable on Gaussian states, such as the logarithmic negativity \cite{logneg,lognegPlenio}). It is particularly interesting to optimise such quantities at steady state, in that in such a regime the filtering operation implies the added advantage of stabilising the state of the open quantum system in the face of noise \cite{mancini06,mancini07,serafozzi10,nurdin12,boundTH}.

\begin{figure}[b!]
\begin{center}
\includegraphics[scale=0.4]{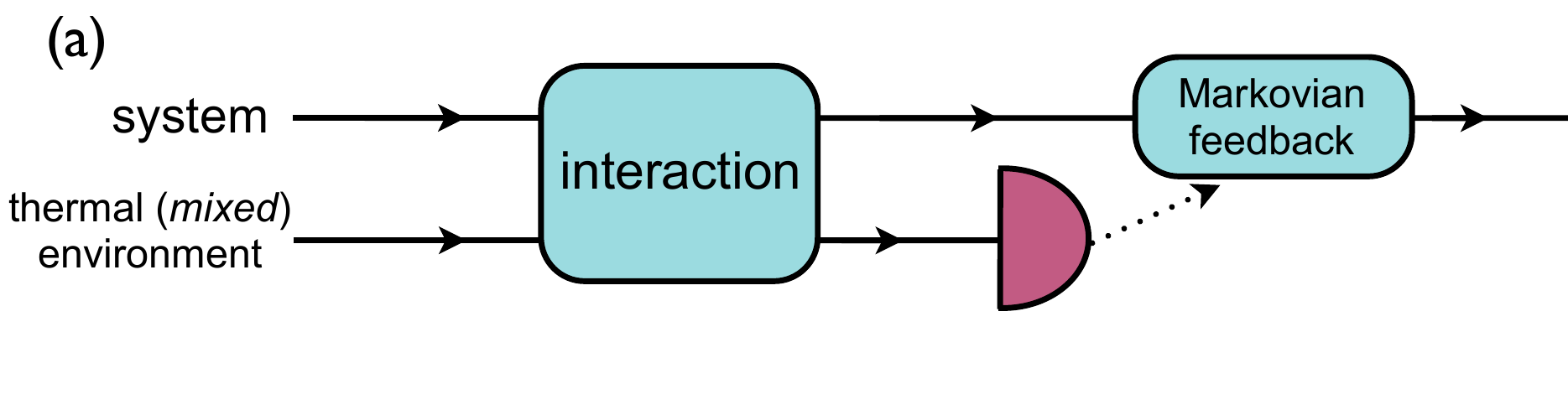}\medskip \\ 
\includegraphics[scale=0.4]{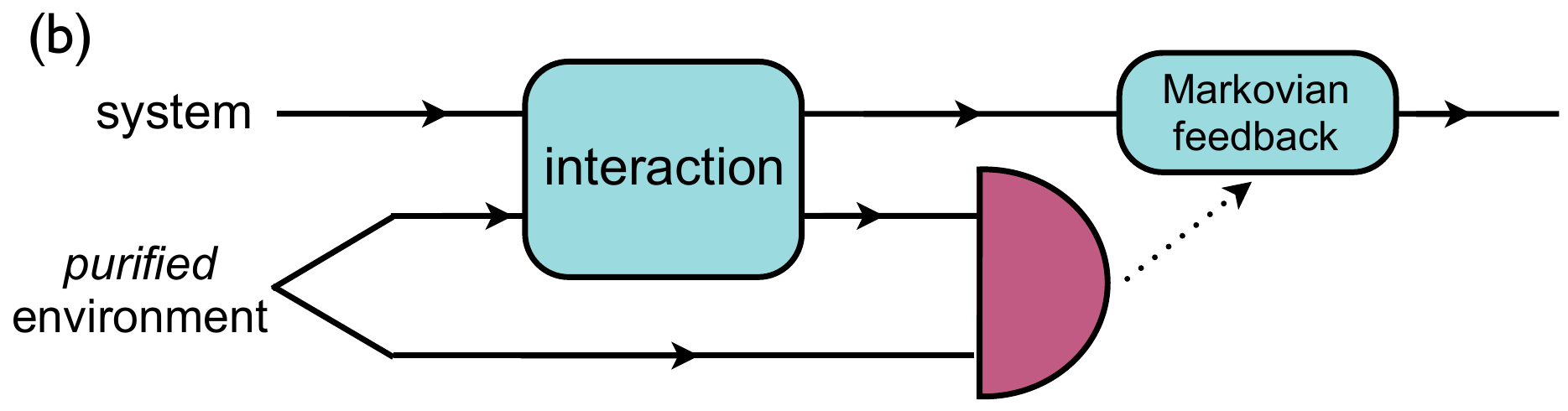}
\end{center}
\caption{Depiction of the configurations considered in the paper. Fig.~1(a) describes the case of total loss ($\gamma=0$), where only the environmental mode interacting with the system, initially in a thermal state, is measured after the interaction. Fig.~1(b)
describes the cases with purified or partially purified environment ($\gamma> 0$), where one measures both the environmental mode interacting with the system and an additional mode entangled with it (in the case $\gamma=1$ the two modes 
are in a global pure state). Access to a purification of the environment allows for larger degrees of asymptotic squeezing.}
\end{figure}

In \cite{boundTH}, 
the optimisation of squeezing -- as well as the closely related one of entanglement in terms of logarithmic negativity -- was addressed for a quantum optical master equation at finite temperature.
There, it was shown that a higher temperature of the bath would in principle lead to higher steady state squeezing under the optimal general-dyne filtering. This result was in part befuddling in that, although 
higher temperature does mean more energy available to achieve higher squeezing, there were strong heuristic 
reasons to believe the monitoring of the environment would not be able to take advantage of an incoherent 
resource, such as thermal noise, to increase a coherent one, such as squeezing along a given phase space 
direction. In particular those results, although concerning with finite temperatures, seemed somewhat at odds with 
the well known no-go theorem of Ref.~\cite{wiseman94a}, establishing that homodyne filtering of a zero temperature 
bath would not allow to generate any squeezing. 

In this paper, we reconsider conditional dynamics resulting from monitoring a bath whose unconditional effect on the system is described by a quantum optical master equation. We define a class of such dynamics that correspond to different incarnations of the monitored bath. At one end [see Fig.~1(a)], one has the continuous homodyne detection of a single-mode bath in a mixed thermal state, leading at steady-state to a thermal state with the same temperature as the environment. At the other end [Fig.~1(b)], one has a bath comprising a thermal mode interacting with the system as well as a second mode that purifies the whole environment, and a filtering based on measuring both modes. In between, one has partial degrees of purification of the bath, parametrised by a nominal loss rate.

We will hence shed light on the apparent inconsistency that higher temperature leads to higher squeezing under the optimal monitoring, by showing that the latter requires the access to a purification of the bath (a global pure state that subsumes the bath thermal state as a local state). More generally, by considering a two-mode bath state resulting from sending the purification above through a lossy channel (with fixed local single-mode states), we will show that any degree of access to a purification of the bath is capable of improving the performance of filtering processes in terms of asymptotic squeezing.
Specifically, the conditional dynamics leads to the maximum amount of steady state squeezing allowed by general-dyne detections only through access to the complete purification ({\em i.e.} in the case of zero loss), . We will also show that a given level of purification of the bath ({\em i.e.} a loss below a certain threshold) is needed in order to observe quantum squeezing for a fixed value of the thermal excitations number.

Notice that our findings may be viewed as a case of entanglement-assistance, 
in the sense that a higher entanglement in the environment modes allows one to improve interesting figures of merit. 
Going beyond the system-bath paradigm, our results are hence relevant to systems where 
several controlled degrees of freedom co-exist in various entangled configurations, such as opto-mechanical set-ups  
where light modes interact with coupled mechanical oscillators \cite{,botter13,seok13}.

\section{Thermal master equation and stochastic unravellings}
We consider here a quantum harmonic oscillator described by bosonic operators $[c,c^\dagger]=\mathbbm{1}$,  interacting 
with a non-zero temperature bath with an average thermal photon number $N$.
The corresponding time evolution is described by the Lindblad Master Equation
\cite{WisemanMilburn}
\begin{align}
d\varrho &= \mathcal{L}_{\sf th} \varrho \: dt \nonumber \\
&= (N+1) \mathcal{D}[c] \varrho \: dt + N \mathcal{D}[c^\dag] \varrho \: dt \label{eq:ME}
\end{align}
where $\mathcal{D}[A]\varrho = A\varrho A^\dag - (A^\dag A \varrho + \varrho A^\dag A)/2$.
We assume to monitor continuously the environment on time scales which are much
shorter than the typical system's response time, by means of
weak measurements. These positive-operator valued measures
(POVMs) are usually referred to as `general-dyne detections' \cite{WisemanMilburn,Belavkin},
and encompass both homodyne and heterodyne detection on the bath degrees of freedom.
The evolution of the conditional state is described by a stochastic master equation (SME)
which is univocally determined by the POVM describing the continuous monitoring.
If the measurement outcomes are not recorded, one has to average over 
all the possible conditional states obtaining an evolution described by the original master equation
(in our case, Eq. (\ref{eq:ME})). For this reason each SME is referred to as an 
`unravelling' of the Lindblad master equation \cite{WisemanMilburn, Carmichael,Gardiner}.

Here, we shall define a class of unravellings of the thermal master equation able to interpolate between two extremal cases. At one extremum, one recovers the SME corresponding to continuous homodyne detection of the bath in a mixed thermal state, leading at steady-state to a thermal state with the same temperature as the environment. At the other extremum, one obtains an unravelling based on accessing and measuring both the environmental mode that interacts with the system and the mode that purifies the latter.

%
%
\section{Generic stochastic unravelling of the thermal master equation}
\label{s:twomodeunravel}
The master equation (\ref{eq:ME}) can be obtained by considering a 
harmonic oscillator interacting by a beam splitting interaction with a bath mode in a 
thermal state at non-zero temperature, with $N$ thermal photons 
on average. Equivalently, one can consider this bath mode being 
part of a two-mode bath, which is in a pure two-mode squeezed 
vacuum state with $N$ average photons in each mode. If we trace 
out the bath mode non-interacting with the system, we indeed 
have a thermal state, and if no information is extracted from the bath, 
the evolution is described by the master equation introduced before. 
For Gaussian states, all the properties we are interested in are determined 
by the covariance matrix, whose elements are defined as 
$\sigma_{jk} = \langle \hat{r}_j \hat{r}_k + \hat{r}_k \hat{r}_j \rangle - 2 \langle \hat{r}_j\rangle \langle \hat{r}_k\rangle$, where $\langle \hat{A}\rangle = \Tr[\varrho \hat{A}]$ and $\hat{{\bf r}}=(\hat{x}_1,\hat{p}_1,\dots,\hat{x}_m,\hat{p}_m)^{\sf T}$ is the vector of quadrature operators for $m$ oscillators.
In the case of a two-mode squeezed vacuum state, the covariance matrix reads
\begin{align}
\sigmaCM_{\sf TMS} = 
\left(
\begin{array}{c | c }
(2 N+1) \mathbbm{1}_2 & 2\sqrt{N(N+1)} \sigma_z \\
\hline 
2\sqrt{N(N+1)} \sigma_z & (2N+1)\mathbbm{1}_2 
\end{array}
\right),
\end{align}
where $\mathbbm{1}_2$ is the two-dimensional identity matrix and $\sigma_z$ is the Pauli $z$ matrix. One can also consider the intermediate case where the 
bath is described by a two-mode state having a covariance matrix
\begin{align}
\sigmaCM_\gamma &= \gamma \sigmaCM_{\sf TMS} + (1-\gamma) \sigmaCM_{\sf Th} \oplus 
\sigmaCM_{\sf Th} \nonumber \\
&= 
\left(
\begin{array}{c | c }
(2 N+1) \mathbbm{1}_2 &  2\gamma\sqrt{N(N+1)} \sigma_z \\
\hline 
2\gamma\sqrt{N(N+1)} \sigma_z & (2N+1)\mathbbm{1}_2 
\end{array}
\right)  \label{eq:CMbath} \:,
\end{align} 
where $ \sigmaCM_{\sf Th}=(2N+1)\mathbbm{1}_2$ represents the covariance matrix of a single-mode thermal state. For any value $0\leq \gamma \leq 1$, the passive interaction between one mode of the bath with the system under exam leads to the thermal master equation (\ref{eq:ME}). 
We notice that Eq. (\ref{eq:CMbath}) can be interpreted as the evolution 
of a two-mode squeezed state in a thermal environment. The parameter $\gamma$ gradually kills the correlations between the two modes of the environment, ranging from a pure, maximally correlated state for 
$\gamma =1$, to two uncorrelated single-mode states for $\gamma=0$. Clearly, in the latter case measurements on the second mode of the bath will not carry any information about the system 
(which only interacts with the first mode).
 
Following this picture, and exploiting the features of the two-mode bath,
we will look for measurement schemes able to interpolate between the two extreme cases
described in the previous section.

Let us consider at time $t$ the quantum state $R(t)=\varrho(t)\otimes \mu(t)$, where 
$\varrho(t)$ and $\mu(t)$ represent respectively the state of the system and 
of the (two-mode) bath.  
In order to describe the effect due to a continuous measurement of the bath, we will
follow the procedure used in Ref. \cite{WisemanThesis}. We start by transforming 
the bath state into a Wigner probability distribution obtaining the operator (in the
system Hilbert space)
\begin{align} 
\widetilde{W}(t) &= \int  \frac{d^2 \lambda_1 d^2\lambda_2 }{\pi^4} \times \nonumber \\
& \times {\rm Tr}_{\sf B} \left[ R(t)\: e^{\left\{\lambda_1 (a^{\dag}-\alpha^*)-\lambda_1^*(a-\alpha) +\lambda_2 (b^{\dag}-\beta^*) - \lambda_2^*(b-\beta) \right\}}\right] \nonumber  \\
&= W_t^{(2)}(\alpha,\beta) \varrho(t)  \label{eq:wigner}
\end{align}
{where we denoted  the Wigner function of a two-mode state having the covariance matrix (\ref{eq:CMbath}) as
\begin{align}
W_t^{(2)}(\alpha,\beta) &= \frac{4}{\pi^2 f(N,\gamma) } \exp\{ -2 (2N+1) (|\alpha|^2 +|\beta|^2) +  \nonumber \\
&\:\: +  8\gamma\sqrt{N(N+1)}
(\alpha_R\beta_R-\alpha_I\beta_I) \}
\end{align}
with $$f(N,\gamma)=\sqrt{{\rm det}[\sigmaCM_\gamma]}=1+4N(N+1)(1-\gamma^2)\:.$$ 
Above in Eq. (\ref{eq:wigner}) we have introduced the bosonic operators $a=\sqrt{dt}e_1(t)$ and
$b=\sqrt{dt} e_2(t)$, satisfying the commutation relations $[a,a^\dag]=[b,b^\dag]=\mathbbm{1}$, 
while the operators $e_j$, which describe the (two-mode) reservoir with infinite bandwidth, 
satisfy $[e_j(t),e_k(t')]=\delta(t-t')\delta_{jk}$.\\}
After an infinitesimal time $dt$ the system and the bath evolve according to the
interaction Hamiltonian $H_{\sf sb} = - i (e_1(t)^\dag c - c^\dag e_1(t))$, such that 
\begin{align}
R(t+dt) &= R(t) + dt [e_1(t)^\dag c - c^\dag e_1(t), R(t)] + O(dt^2) \\
&= R(t) + \sqrt{dt} [ a^\dag c-c^\dag a, R(t)] + O(dt) \:,
\end{align}
which in the Wigner function picture reads
\begin{align}
\widetilde{W}(t+dt) &= \widetilde{W}(t) + \sqrt{dt}\left[ (\alpha^* -\frac12 \partial_\alpha) c \widetilde{W}(t) - \right. \nonumber \\
& \: \left. - (\alpha+\frac12 \partial_{\alpha^*})c^\dag \widetilde{W}(t ) - (\alpha^*+\frac12\partial_\alpha) \widetilde{W}(t) c + 
\right. \nonumber \\ 
&\:\: + \left. (\alpha-\frac12 \partial_{\alpha^*}) \widetilde{W}(t) c^\dag \right] + O(dt) \label{eq:wignerevolv} \:.
\end{align}
We now consider the simultaneous measurement of the quadratures $\hat{q}_A = a+a^\dag$ and $\hat{q}_B=b+b^\dag$, corresponding to the two different modes characterizing the bath. 
Notice that a more general Gaussian measurement might be considered, such as the {\em general-dyne} measurement of the operators $a+\Upsilon_a a^\dag$ and $b+\Upsilon_b b^\dag$, with $-1\leq \Upsilon_{a,b}\leq 1$ (see, e.g., \cite{GeneralDino}). However, in this case simple homodyne measurements will prove apt to get the results desired.
The (unnormalized) conditional state of the system can be obtained through the equation
\begin{align}
\widetilde{W}_c (t+dt) = \int d^2 \alpha \: d^2 \beta \: \widetilde{W}(t+dt) \delta(2 \alpha_R -q_A) 
\delta(2 \beta_R -q_B). 
\end{align}
By performing the derivatives and the integrals one obtains
\begin{widetext}
\begin{align}
\widetilde{W}_c(t+dt) = p(q_A,q_B; t) &\left\{
\varrho(t) + \left[ h_1(N,\gamma) (N+1) (c\varrho(t) +\varrho(t) c^\dag) + h_2(N,\gamma) N (c^\dag \varrho(t) +\varrho(t) c)\right] \frac{\sqrt{dt} \: q_A}{f(N,\gamma)}
\right. \nonumber \\
& \left. -\gamma \sqrt{N (N+1)}\left[c\varrho(t) + \varrho(t) c^\dag + c^\dag \varrho(t) + \varrho(t) c\right] 
 \frac{\sqrt{dt}\: q_B}{f(N,\gamma)}
\right\} + O(dt) \:, \label{eq:unravel1}
\end{align}
\end{widetext}
where the probability of measuring $q_A$ and $q_B$ at time $t$ $p(q_A,q_B ; t)$ is a two-variable Gaussian distribution centered in zero and having a covariance matrix
\begin{align}
{\bf C} = \left(
\begin{array}{c c}
2N+1 & 2 \gamma \sqrt{N (N+1)} \\
2\gamma \sqrt{N (N+1)} & 2 N+1 
\end{array}
\right) \:, \label{eq:CM}
\end{align}
and we have defined the functions
\begin{align}
h_1(N,\gamma) &= 1+2N(1-\gamma^2) \\
h_2(N,\gamma) &= 2\gamma^2 -1 +2N(\gamma^2-1).
\end{align}
Consequently the probability of measuring $q_A$ and $q_B$ at time $t +dt$ can be calculated, obtaining
\begin{widetext}
\begin{align}
p(q_A,q_B ; t+dt) &= \hbox{Tr}_s [ \widetilde{W}_c(t+dt) ] \\
&= p(q_A,q_B; t)  \left\{ 1+ \frac{2N+1}{f(N,\gamma)}\langle c+c^\dag \rangle  \sqrt{dt} \: q_A  -2\frac{\sqrt{N(N+1)}}{f(N,\gamma)} \langle c+c^\dag \rangle \sqrt{dt} \: q_B\right\} + O(dt) \:.
\end{align}
\end{widetext}
We can thus write the two quadratures of the bath as Gaussian random variables
satisfying
\begin{align}
\sqrt{dt} \:q_A &= \langle c+c^\dag \rangle dt + dw_A \label{eq:xA} \\
\sqrt{dt} \:q_B &= dw_B \:, \label{eq:xB}
\end{align}
where we defined the correlated Wiener increments s.t.
\begin{align}
\left(
\begin{array}{c c}
dw_A^2 & dw_A dw_B \\
dw_A dw_B & dw_B^2 
\end{array}
\right) = {\bf C} \:dt \:,
\end{align}
with the covariance matrix ${\bf C}$ defined in Eq. (\ref{eq:CM}). 
Hence the normalized conditional state can be written as 
\begin{widetext}
\begin{align}
\varrho_c (t+dt) &= \frac{\widetilde{W}_c(t+dt)}{p(q_A,q_B;t+dt)} \\
&= \left\{ 1 + \mathcal{H}[h_1(N,\gamma)(N+1) c + h_2(N,\gamma) N c^\dag] \frac{\sqrt{dt}\: q_A}{f(N,\gamma)} - \gamma \mathcal{H}[\sqrt{N (N+1)}(c+c^\dag) ] \frac{\sqrt{dt} \: q_B}{f(N,\gamma)}
\right\} \varrho(t) + O(dt) , \label{eq:normalised}
\end{align}
\end{widetext}
where $\mathcal{H}[A]\varrho = A\varrho + \varrho A^\dag - \hbox{Tr}[(A+A^\dag)\varrho]\varrho$. 
Substituting $q_A$ and $q_B$ by using Eqs. (\ref{eq:xA}) and (\ref{eq:xB}) one obtains
the stochastic master equation (SME) corresponding to this measurement scheme
\begin{widetext}
\begin{align}
d\varrho_c (t+dt) &= \varrho_c (t+dt) -\varrho_c(t) \nonumber \\
&= \mathcal{H}[h_1(N,\gamma) (N+1) c + h_2(N,\gamma) Nc^\dag] \varrho_c(t) \frac{dw_A}{f(N,\gamma)}  -\mathcal{H}[\gamma \sqrt{N (N+1)}(c+c^\dag) ] \varrho_c(t) \frac{dw_B}{f(N,\gamma)} 
+ O(dt) \:. \label{eq:SME1}
\end{align}
\end{widetext}
We remark that, as the average of all the terms we have explicitly derived is zero, the $O(dt)$ term must equal the RHS of Eq. (\ref{eq:ME}).
One can now recast the obtained SME (\ref{eq:SME1}) in terms of uncorrelated Wiener increments
\begin{align}
\left(
\begin{array}{c}
d\widetilde{w}_A \\
d\widetilde{w}_B
\end{array}
\right) =
{\bf M}^{-1} \left(
\begin{array}{c}
dw_A \\
dw_B
\end{array}
\right)    
\end{align}
where the matrix
\begin{align}
{\bf M} = \left(
\begin{array}{c c}
m_+(N,\gamma) &  m_-(N,\gamma) \\
m_-(N,\gamma) & m_+(N,\gamma)
\end{array}
\right) \:,
\end{align}
with
\begin{align}
m_{\pm}(N,\gamma)&=\sqrt{\frac{1+2N\pm \sqrt{f(N,\gamma)}}{2}} \:,
\end{align}
is such that ${\bf C}={\bf M}{\bf M}^{\sf T}$, and, as said above
\begin{align}
\left(
\begin{array}{c c}
d\widetilde{w}_A^2 & d\widetilde{w}_A d\widetilde{w}_B \\
d\widetilde{w}_A d\widetilde{w}_B & d\widetilde{w}_B^2 
\end{array}
\right) = \mathbbm{1} \:dt\:.
\end{align}
By substituting the previous Wiener increments with the new ones, we finally obtain
\begin{widetext}
\begin{align}
d\varrho_c (t+dt) &= \frac{m_+(N,\gamma)}{f(N,\gamma)} \mathcal{H}[h_1(N,\gamma)(N+1)c + 
h_2(N,\gamma) N c^\dagger ] \varrho_c(t)d \widetilde{w}_A -
\frac{ m_-(N,\gamma)}{f(N,\gamma)} \mathcal{H}[\gamma\sqrt{N(N+1)} (c + c^\dagger) ] \varrho_c(t) d\widetilde{w}_A + \nonumber \\
&+\frac{m_-(N,\gamma)}{f(N,\gamma)} \mathcal{H}[h_1(N,\gamma)(N+1)c + 
h_2(N,\gamma) N c^\dagger ] \varrho_c(t)d \widetilde{w}_B -
\frac{ m_+(N,\gamma)}{f(N,\gamma)} \mathcal{H}[\gamma\sqrt{N(N+1)} (c + c^\dagger) ] \varrho_c(t) d\widetilde{w}_B + \nonumber \\
&\:\: + O(dt) \:. \label{eq:SME2}
\end{align} 
\end{widetext}
If we set $\gamma=0$, that is by considering the bath is in an uncorrelated thermal state, the monitoring of the ancillary mode $b$ cannot influence the system's evolution, and the 
corresponding SME reads
\begin{align}
d\varrho_c(t) &=  \mathcal{H}[(N+1) c - N c^\dagger] \varrho_c(t) \frac{d\widetilde{w}_A}{\sqrt{(1+2N)}}  + O(dt) \:.
\end{align}
which is the well known unravelling due to continuous homodyne detection on the mixed thermal bath
\cite{WisemanMilburn,WisemanThesis,GeneralDino}.
On the other hand for $\gamma=1$, that is when the two-mode bath is in a maximally entangled two-mode squeezed vacuum state, we have
\begin{align}
d\varrho_c (t+dt) 
&= \sqrt{N+1}\:\mathcal{H}[c] \varrho(t)\: d\widetilde{w}_A + \nonumber \\
& \:\:\: - \sqrt{N}\:\mathcal{H}[c^\dag ] \varrho(t)\: d\widetilde{w}_B + O(dt) \:. \label{eq:SMEopt}
\end{align}
This indeed corresponds to the optimal unravelling proposed in \cite{boundTH}, 
achieving at steady state the maximum amount of squeezing in the quadrature $\hat{X} = c+c^\dag$.
Such squeezing saturates the inequality
\begin{equation}
V_x \equiv \langle\hat{X}^2 \rangle -\langle \hat{X}\rangle^2 \ge \frac{1}{2N+1} \; ,  \label{bound}
\end{equation}
which holds at steady state for all unravellings of Eq.~(\ref{eq:ME}).
Note that, here, we refer to maximum achievable squeezing as to the minimum variance of a quadrature operator.

\section{Analysis of the purified stochastic unravellings}

To fully understand the properties of the  general SME (\ref{eq:SME2}) in terms of the free parameter $\gamma$, let us now study the behaviour of the {\em indirectly} monitored quadrature $\hat{X}$, studying in particular its variance at steady state.
We recall that in the presence of the master equation (\ref{eq:ME}), the evolution of the average value $\langle\hat{X}\rangle$ and its variance $V_x = \langle \hat{X}^2\rangle - \langle \hat{X}\rangle^2$ read
\begin{align}
d\langle \hat{X} \rangle &= - \frac{\langle \hat{X} \rangle}{2} dt  \\
dV_x &= (2 N+1 - V_x ) dt \:. 
\end{align}
If we rather consider the SME (\ref{eq:SME2}), by using the formulas
\begin{align}
\hbox{Tr} [ \mathcal{H}[c] \varrho (c+c^\dag)] = V_x -1 \:, \\
\hbox{Tr} [ \mathcal{H}[c^\dag] \varrho (c+c^\dag)] = V_x +1\:,
\end{align}
we obtain
\begin{align}
d\langle X \rangle &= - \frac{\langle X \rangle}{2} dt  +[ A_1(N,\gamma) V_x + A_2(N,\gamma) ]\:d\widetilde{w}_A + \nonumber \\
& \: + [B_1(N,\gamma)V_x + B_2(N,\gamma) ]\: d\widetilde{w}_B\:, 
\end{align}
where
\begin{widetext}
\begin{align}
A_1(N,\gamma)&=\frac{m_+(N,\gamma) [(N+1) h_1(N,\gamma) + N h_2(N,\gamma)]-2\gamma
m_-(N,\gamma)\sqrt{N(N+1)}}{f(N,\gamma)} \nonumber \\
A_2(N,\gamma)&=\frac{m_+(N,\gamma) [(N+1) h_1(N,\gamma) - N h_2(N,\gamma)]}{f(N,\gamma) }\nonumber \\
B_1(N,\gamma)&=\frac{m_-(N,\gamma) [(N+1) h_1(N,\gamma) + N h_2(N,\gamma)]-2\gamma
m_+(N,\gamma)\sqrt{N(N+1)}}{f(N,\gamma)}\nonumber \\
B_2(N,\gamma)&=\frac{m_-(N,\gamma) [(N+1) h_1(N,\gamma) - N h_2(N,\gamma)]}{f(N,\gamma)} \:.\nonumber
\end{align}
\end{widetext}
Consequently, by considering an input Gaussian state and by using Ito calculus, 
the evolution equation for the variance reads
\begin{align}
dV_x &= d\langle X^2 \rangle - 2 \langle X \rangle d\langle X\rangle - (d\langle X \rangle )^2 \\
&=  \left[ 2N+1 - V_x - (A_1(N,\gamma) V_x + A_2(N,\gamma))^2 + \right. \nonumber \\
& \: \left. - (B_1(N,\gamma) V_x + B_2(N,\gamma))^2 \right] dt \:.
\end{align}
\begin{figure}[b!]
\begin{center}
\includegraphics[width=0.9\columnwidth]{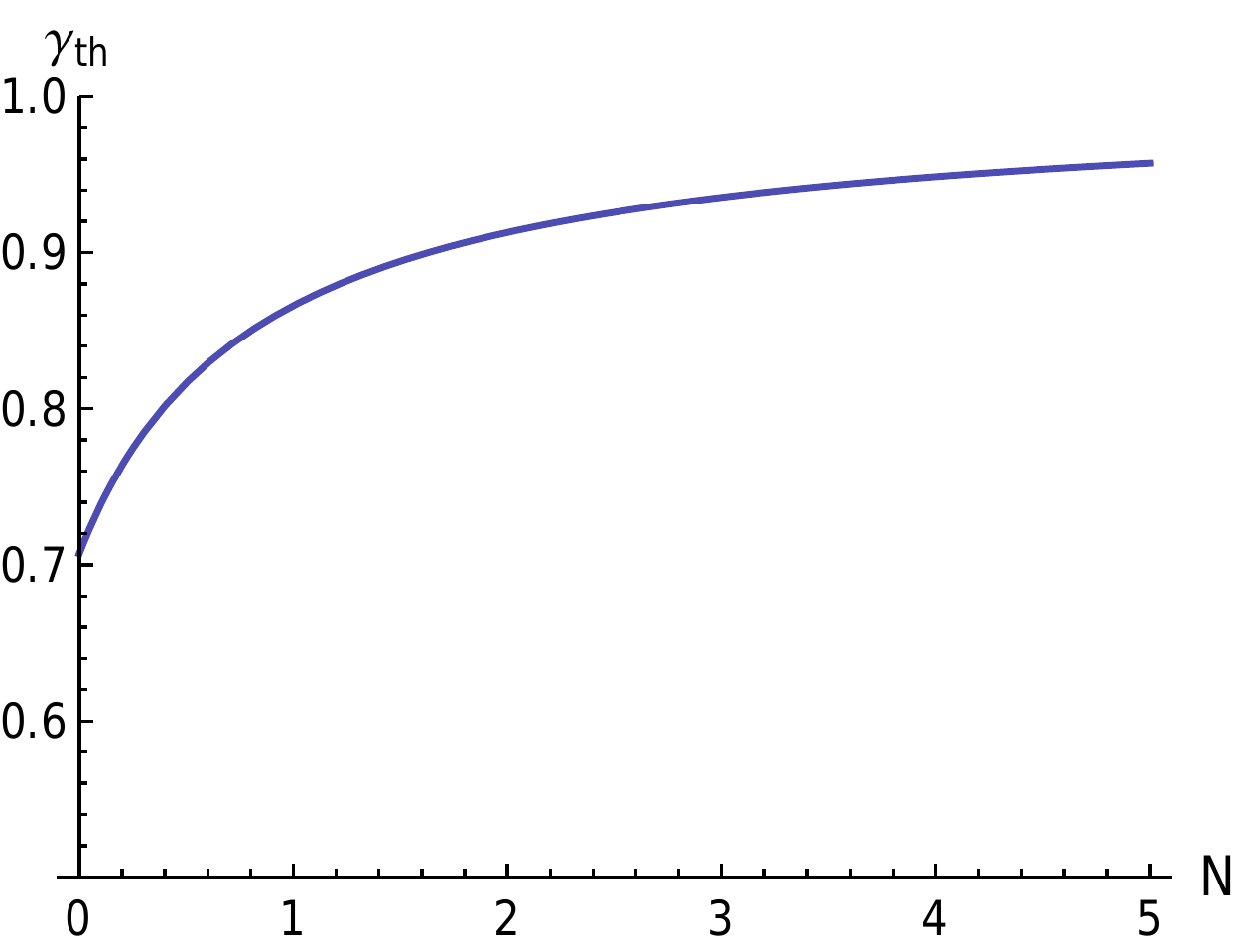}\\ 
\end{center}
\caption{Threshold value for the purification parameter $\gamma_{\sf th}(N)$ needed to observe quantum squeezing, as a function of the number of thermal excitations $N$. 
\label{f:gammath}}
\end{figure}
Notice that for Gaussian states, the stochastic master equation (\ref{eq:SME2}) gives a deterministic evolution of the second moments \cite{wiseman05}, and no Wiener increments are present in the corresponding equation.
By posing the steady-state condition  $dV_x/ dt =0$, we obtain as the only phyiscal solution
\begin{align}
V_x^{\sf (ss)} = 2N+1 -\gamma^2 \frac{4 N(N+1)}{2N+1} \:. \label{eq:steadyvariance}
\end{align}
As expected, varying the parameter $\gamma$, $V_x^{\sf (ss)}$ decreases monotonically from the thermal variance $V_x=2N+1$ to the optimal squeezed variance $V_x=1/(2N+1)$, which saturates the bound derived in \cite{boundTH} 
(notice that it must be optimal, as a diagonal entry of a matrix must be larger than or equal to the smallest eigenvalue). 
Moreover we observe that, in order to obtain quantum squeezing, {\em i.e.} quadrature fluctuations below the vacuum level, one needs a value of $\gamma$ above the threshold
\begin{align}
\gamma > \gamma_{\sf th} (N)= 
\sqrt{\frac{2 N+1}{2 (N+1)}} .
\end{align}
\begin{figure}[b!]
\begin{center}
\includegraphics[width=0.9\columnwidth]{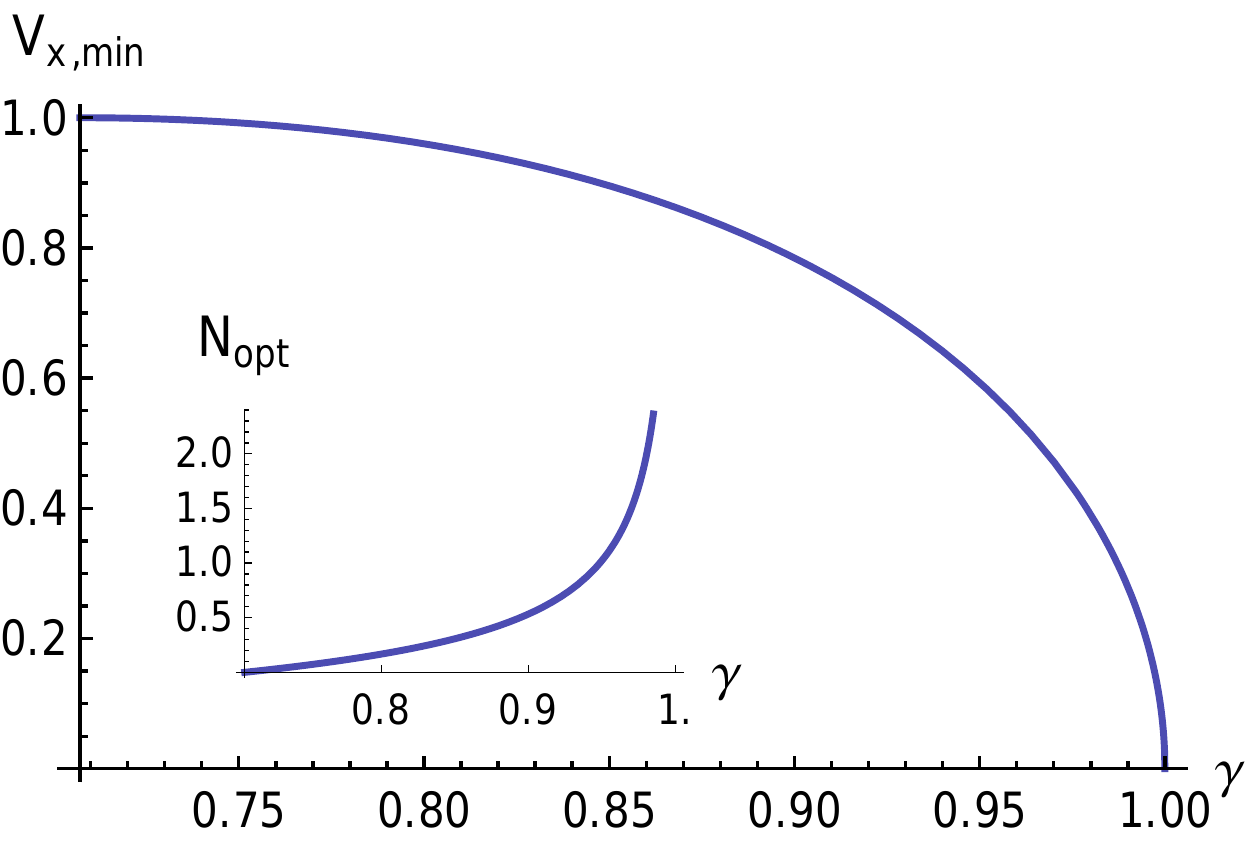}\\ 
\end{center}
\caption{Minimum variance $V_{x,{\sf min}}$ achievable for fixed value of the purification parameter $\gamma$. Inset: optimal value of the thermal excitations number $N_{\sf opt}$ minimizing the variance $V_x$ for a given value of $\gamma$.  
\label{f:Vxmin}}
\end{figure}
As one can observe from Fig. \ref{f:gammath}, $\gamma_{\sf th}(N)$ has as minimum value $\gamma=1/\sqrt{2}$ and is an increasing function of the number of thermal excitations $N$, showing how a less noisy purification of the bath is needed at higher temperatures.
If we rather fix the value of $\gamma$, we find that for $\gamma<1/\sqrt{2}$ the variance $V_x$ is an increasing function of the number of thermal excitations $N$, such that the minimum value achievable is $V_{x,{\sf min}}=1$ and, as mentioned above, no squeezing can be obtained. On the other hand for a larger loss factor, $\gamma>1/\sqrt{2}$, a minimum in the variance is observed for a certain $N=N_{\sf opt}$: Such and optimal (minimum) value of the achievable variance reads
\begin{align}
V_{x,{\sf min}} = 2 \gamma \sqrt{1-\gamma^2}.
\end{align}
Its behaviour is reported in Fig. \ref{f:Vxmin}, where we observe that it decreases monotonically to zero with $\gamma$. Analogously, in the inset we plot the behaviour of $N_{\sf opt}$, which increases monotonically to infinity as $\gamma$ tends to one. 

The same procedure can be used in order to derive the evolution equation for the variance of the conjugated quadrature $V_{p}$ and for the covariance $V_{xp}$. One then obtains the steady-state values $V_p^{\sf (ss)} =2N+1$, and $V_{xp}^{\sf (ss)}=0$ for all values of $\gamma$, which correspond to the case without continuous measurement. We can calculate the purity $\mu^{\sf (ss)}={\rm Tr}\varrho^{\sf (ss) 2}$ of the steady-state Gaussian state
$\varrho^{\sf (ss)}$, obtaining
\begin{align}
\mu^{\sf (ss)} = \frac{1}{\sqrt{1+4N(N+1)(1-\gamma^2)}}, 
\end{align}
which is equal to unit only for $N=0$ ($\forall \gamma$) or $\gamma=1$ ($\forall N$). Let us briefly remind the reader that the purity of a single-mode Gaussian state $\varrho$ with covariance matrix $\sig$ is given by 
$\mu=1/\sqrt{{\rm Det}\sig}$.

The derived SME is able now to shed light on the counterintuitve claim that higher temperature of the bath can in principle lead to a larger amount of squeezing at steady-state. It is now clear that, in order to achieve the aforementioned bound, one needs to be able to access and monitor the purification of the bath (in this case, the two-mode squeezed vacuum state). In particular, in order to obtain a steady-state variance smaller than the thermal one, $V_x=2N+1$, one needs a value of $\gamma>0$, that is one needs to access information on the ancillary bath mode, while an even larger value $\gamma > \gamma_{\sf th}(N)$ is needed in order to observe quantum squeezing.
\section{Conclusions}
Our study emphasizes that, when the environment is in a thermal state at non-zero temperature, the class of the possible unravellings comprises the possibility to monitor a (complete or partial) purification of the environment. In particular the maximal asymptotic squeezing allowed by general-dyne detection is only achievable when a complete purification is accessible.
As immediately clear from the treatment 
of Ref.~\cite{boundTH}, the same argument applies to the asymptotic Gaussian entanglement.
More generally, any level of purification will help with 
respect to what is allowed by accessing the mode interacting with the system alone (which, in the 
absence of interactions for the system, does not allow for any squeezing at all, in line with 
what was already known at zero temperature \cite{wiseman94a}). 
Furthermore, a threshold value on the purification parameter is derived such that, at a given temperature, 
quantum squeezing of the bath can be observed only for $\gamma>\gamma_{\sf th}(N)$.

Our results shed considerable light on the control possibilities allowed by quantum filtering through Gaussian measurements. Particularly they put forward the idea of entanglement assisted feedback control and they might be also of interest to systems where 
a few canonical degrees of freedom are under control, several of which can be simultaneously monitored. 
This could be the case, for instance, in quantum opto-mechanics where the interactions of several light 
and mechanical modes can in principle be brought under control.

\acknowledgments AS and MGG acknowledge support from EPSRC through grant EP/K026267/1.
HMW acknowledges support by the Australian Research Council Centre of Excellence CE110001027.

\end{document}